# Direct Radiative Impacts of Stratospheric Aerosols on the Tropical Troposphere: Clouds, Precipitation, and Circulation in Convection-Resolving and Global Simulations

Zachary McGraw[1,2] and Lorenzo M. Polvani[1,3,4]

*1 Department of Applied Physics and Applied Mathematics, Columbia University, New York, NY*

*2 NASA Goddard Institute for Space Studies, New York, NY, USA*

*3 Department of Earth and Environmental Sciences, Columbia University, New York, NY*

*4 Division of Ocean and Climate Physics, Lamont-Doherty Earth Observatory, Palisades, NY*

Corresponding e-mail: zachary.mcgraw@columbia.edu

ABSTRACT

A concern for stratospheric aerosol injection (SAI) is that stratospheric aerosols could inadvertently alter rain and winds through mechanisms independent of the intended surface cooling. We here use a multi-model framework to investigate how the tropical troposphere responds to SAI when sea surface temperatures are held fixed. By performing convection-resolving simulations in small-domains and in mock-Walker setups, and contrasting these with global climate model simulations, we trace how stratospheric aerosols radiatively heat the troposphere, and in turn alter convection, clouds, and rainfall. Our simulations show an SAI-induced reduction in tropical mean precipitation, yet decreased cloud radiative heating moderates this effect and complicates its predictability. Regional rainfall anomalies within the tropics can be substantial. However, surface-temperature-independent effects on tropical circulation are found to be negligible, indicating that stratospheric aerosols do not inherently alter the tropical overturning circulation as previously suggested. These results clarify the mechanisms governing SAI hydroclimate impacts and show that key uncertainties arise from cloud processes that models are unable to constrain. Consequently, near-term SAI deployment would carry the risk of being implemented without the ability to reliably predict its hydroclimate impacts.

SIGNIFICANCE STATEMENT

Stratospheric aerosol injection could help stabilize Earth's temperature, but models have struggled to reliably predict its unintended climate impacts. This study uses a multi-model framework, including a novel application of convection-resolving simulations, to examine responses of clouds and rainfall to stratospheric aerosol injection in depth. Our findings clarify the drivers and limits of stratospheric aerosol impacts on tropical precipitation through mechanisms independent of surface cooling. Especially highlighted is how cloud radiative responses drive intermodel differences and reduce confidence in projections. Contrary to previous suggestions, our simulations show no indication of weakened tropical circulation via surface-temperature-independent processes. These findings offer key insights for efforts to predict the hydroclimate risks associated with deliberate solar radiation management.

1. Introduction

While stratospheric aerosol injection (SAI) could potentially counteract the impact of greenhouse gas, there is concern that stratospheric aerosols might inadvertently alter underlying tropospheric conditions and hydroclimate. SAI is a climate intervention strategy that involves releasing reflective aerosols, typically sulfuric acid particles, into the stratosphere to cool Earth's surface by reducing incoming sunlight (Crutzen, 2006; National Academies, 2021). Yet these aerosols also perturb the atmosphere through mechanisms not directly tied to surface temperature change. By absorbing, emitting, and scattering radiation, stratospheric aerosols alter the atmospheric energy balance and vertical structure, producing what are often termed "fast" or "rapid" responses (Andrew et al., 2010; Bala et al., 2010) in contrast with slower adjustments mediated by sea surface temperatures (SSTs). In this study we refer to the former as *SST-independent*, or *surface-temperature-independent* in cases that can be generalized because land cooling is absent or does not substantially affect the conclusions. Among SAI risks, surface-temperature-independent responses may be the most critical to constrain, as they represent side-effects inherent to the intervention, rather than intended changes aimed at counteracting greenhouse gas forcing.

Global modeling studies have provided a range of evidence linking stratospheric aerosols to tropospheric anomalies via mechanisms independent of surface cooling. Using an intermediate-complexity global climate model (GCM), Ferraro et al. (2014, 2016) argued that these effects weaken tropical circulation and suppress mean rainfall. Usha et al. (2024) found comparable rainfall reductions in a full GCM and showed that the magnitude of these reductions is sensitive to aerosol layer height. Simpson et al. (2019) isolated the effect of SAI-induced stratospheric heating in a GCM, confirming its role in precipitation suppression. Similarly, McGraw and Polvani (2024) and Raiter et al. (2025) reported substantial SST-independent precipitation reduction following large volcanic eruptions in GCM simulations.

Likewise, the IPCC AR6 has highlighted the fact that even if climate intervention were to successfully offset global warming, substantial residual and regionally variable climate anomalies would be likely. These responses could potentially have societally significant consequences, even under stabilized global mean temperatures. Compounding this concern, the precipitation response to SAI has been found to vary widely across global models (Ferraro et al., 2016; Laakso et al., 2024), emphasizing the need for better understanding and tighter constraints on these effects. Surface-temperature-independent responses therefore remain a critical source of uncertainty when evaluating the risks and feasibility of SAI.

Here, we investigate the physical mechanisms driving surface-temperature-independent responses to stratospheric aerosols in the tropics by performing and analyzing limited-area, cloud-resolving simulations and global climate model simulations. This multi-scale framework allows us to isolate and explore the interactions among radiative fluxes, convection, clouds, and precipitation. Our primary goals are to advance understanding of tropospheric responses to SAI and to identify the factors that must be addressed to reliably predict associated hydrological responses. Additionally, we demonstrate that tropical circulation responses are less inherent to stratospheric aerosols than has previously been suggested, as no surface-temperature-independent circulation response is found to exist.

## 2. Methods

*a) Limited-domain convection-resolving simulations*

To examine fast responses to stratospheric sulfate in controlled simulations that explicitly resolve convection, we employ the System for Atmospheric Modeling (SAM; Khairoutdinov & Randall, 2003) version 6.11.9. All simulations are run in radiative–convective equilibrium (RCE) mode, in which a patch of tropical ocean is represented as a three-dimensional domain with doubly-periodic boundaries. Doubly-periodic models are most commonly configured with zero Coriolis force, motivating our focus on the tropics for this study. For all SAM simulations, we use the Predicted Particle Properties (P3; Morrison and Milbrandt, 2015) microphysics scheme together with the Rapid Radiative Transfer Model (RRTM; Iacono et al., 2008) radiation scheme.

We use three setups: (1) a small domain with 1 km horizontal resolution, (2) a small domain with 3 km resolution, and (3) a mock-Walker configuration with 3 km resolution. All these configurations are visually depicted in Fig. 1, which gives a sense of the degree of cloud organization in each setup. The small-domain SAM cases use a 128 × 128 horizontal grid with 84 vertical levels, following the RCEMIP phase 1 (RCEMIP1; Wing et al., 2018a) design but extending the model top from 33 to 38 km given our focus on stratospheric aerosols. The mock-Walker setup is based on RCEMIP phase 2 (Wing et al., 2024), with sinusoidal SST gradients across a 6144 km × 384 km domain having 3 km resolution; here too we use 84 vertical levels. The imposed SST gradient is 5°C, following Dagan et al. (2023), who identify this range as representative of tropical variability. The mean SST is set at 300 K in all simulations. Each experiment is integrated for at least 300 days, and we generally present averages over days 151-300. For the mock-Walker runs, which can exhibit weeks-long disequilibria, we adjust this averaging window to capture equilibrium climate states on average: days 131-280 for the base SAI case, days 141-290 for the small-particle sensitivity test, and days 166-315 for an aerosol experiment designed like the base case but including only aerosol longwave effects (see Section 2c for a more detailed description).

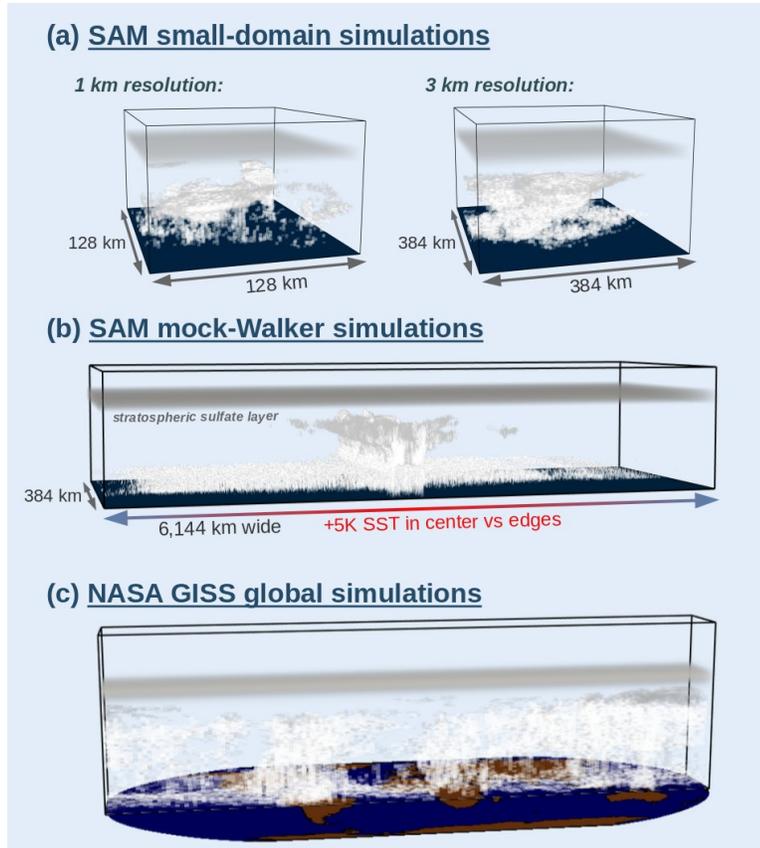

FIG. 1. Instantaneous cloud occurrence in our various model configurations. For SAM, snapshots are shown after 300 days for the small-domain simulations and after 100 days for the mock-Walker simulation (day 300 exhibits irregularities). For GISS, we show clouds only in the tropics (20°S–20°N) to match our focus area, visualized at noon on the day following the 12-year simulation period. Grey layers above the clouds indicate the presence of stratospheric aerosols, which were applied similarly in all simulations.

*b) Global simulations*

To capture the full influence of large-scale dynamics and understand the responses to SAI in a comprehensive model, we also employ the high-top NASA GISS ModelE2.2 global climate model (Orbe et al., 2020). Versions of the GISS model have been used for SAI experiments within the Geoengineering Model Intercomparison Project (GeoMIP; e.g., Visioni et al., 2023), and also to study volcanic sulfate impacts (e.g., Hansen et al., 1992; DallaSanta & Polvani, 2022; McGraw et al., 2024). The E2 series has a horizontal resolution of 2° latitude × 2.5° longitude, with E2.2 extending to 102 vertical levels up to 89 km. Following McGraw and Polvani (2024), which used simulations produced by DallaSanta and Polvani (2022), we prescribe SSTs to values from a control run without stratospheric aerosols, ensuring no ocean surface cooling occurs in response to the aerosols, although land temperatures are still able to adjust. Atmospheric $CO_2$ is set to its mean concentration in the year 2000. Each experiment is run for 12 years to reduce internal variability. For comparison with our SAM RCE simulations, we focus on the tropics between 20°S and 20°N, as is illustrated in Fig. 1c.

*c) Stratospheric aerosol forcings*

In all our stratospheric aerosol injection (SAI) experiments, we impose a stratospheric sulfuric acid layer having an aerosol optical depth (SAOD) of 0.35. This value is near the model-median SAOD that the GeoMIP G6sulfur experiment estimates is needed to maintain an adequately stable surface temperature at the year 2100 (Visioni et al., 2024), and is on the high end of typical SAI scenarios. Whereas G6sulfur calculates meridionally varying SAOD interactively following near-equatorial $SO_2$ injections, we here simply prescribe a uniform SAOD of 0.35 in all columns: given the absence of an accepted latitude-dependent injection strategy, we opt for the simplest distribution. In our primary SAI experiment (hereafter, the *base SAI* case), we set aerosol particle size to have an effective radius of 0.70 μm at all locations. In a sensitivity experiment (*small aerosols* case), we instead use 0.35 μm with the same SAOD, which requires only half the sulfur mass as the base SAI case due to the smaller aerosols generating a stronger forcing per mass (Lacis, 2015). These two sizes roughly correspond to (1) a scenario in which aerosol size evolves naturally from injected $SO_2$ and (2) a hypothetical case in which small condensed particles are directly injected to maximize the cooling-per-mass efficacy (Weisenstein et al., 2022).

In our base and small aerosols experiments with the GISS model, we center the aerosol layer at a mean altitude of 23 km at the equator, decreasing sinusoidally to 18 km at the poles, with a Gaussian vertical distribution having a standard deviation of 2 km. In a sensitivity experiment (*low aerosols* case), the layer is lowered to 18 km at the equator and 12 km at the poles, with a narrower vertical spread of 0.8 km. The SAM simulations mirror these distributions while being representative of the tropics, with identical aerosols prescribed in each column. Accordingly, the distributions are centered at 22.5 km in the base and small aerosols cases, and at 17.5 km in the low aerosols case. All SAM experiments use the same SAOD, effective radius, and vertical distributions as the corresponding GISS simulations.

As aerosols are not a standard feature in SAM, we implemented the radiative effects of a stratospheric sulfate layer directly into SAM's RRTM radiation scheme by specifying appropriate extinction, asymmetry factor, and single-scattering albedo values. To maintain consistency with the GISS experiments described above, we replicated the stratospheric sulfate aerosol optical properties used in the GISS model. This approach assumes a gamma size distribution for the aerosols and employs refractive indices from Palmer and Williams (1975), choices that have been shown to reproduce volcanic aerosol observations (Hansen et al., 1996). In a final sensitivity experiment, we simulate stratospheric aerosols as in our base SAI case but without sulfate effects on shortwave radiation, leaving only longwave (LW) effects (hereafter the *LW only* case).

*d) Vertically-resolved atmospheric energy diagnostics*

To quantify and understand the impact of stratospheric sulfate on atmospheric radiative cooling, we implemented custom vertically-resolved radiative flux diagnostics in both models. These diagnostics recompute radiation at each timestep while selectively omitting clouds and/or aerosols. While all-sky fluxes and clear-sky (cloud-free) cases are already standard, we added clean-sky (aerosol-free) and clean-clear-sky (neither clouds nor aerosols) diagnostics. In GISS, the clean-sky calculations specifically exclude stratospheric sulfate while retaining background aerosols, whereas in SAM, no tropospheric aerosols are present. Using these diagnostics, we decompose the sulfate fast response in radiative heating ($\Delta R$) as follows, with positive terms denoting energy gain by the atmosphere:

$$\Delta R = ARH + RA_{cloud} + RA_{clear-sky} \qquad \text{Eqn. 1}$$

Here, the aerosol radiative heating (ARH) is defined as the difference between all-sky and clean-sky radiative fluxes, and is alternatively referred to as the aerosol *forcing* to highlight its role as the driving perturbation – technically known as an instantaneous radiative forcing (IRF). We define cloud radiative adjustments ($RA_{cloud}$) as the difference between clean-sky and clean-clear-sky radiative flux responses, extending the recommendation of Ghan et al (2013) to vertically-resolved diagnostics. Clear-sky radiative adjustments ($RA_{clear-sky}$) are then treated as the residual relative to the net radiative heating response, $\Delta R$.

In addition to quantifying the radiative flux responses to SAI, we also examine how these responses interact with other terms in the atmospheric energy budget. Over a sufficiently long averaging period for the climate system to re-equilibrate, radiative heating anomalies induced by SAI over a large region are roughly balanced by changes in latent and sensible heating, yielding the energetic constraint (O'Gorman et al., 2012):

$$\Delta R + L\Delta P + \Delta S \approx 0 \qquad \text{Eqn. 2}$$

Here, the latent heating term reflects the change in surface precipitation, ΔP, multiplied by the latent heat of condensation, L. The sensible heating term, ΔS, represents changes in turbulent heat exchange between the surface and the lowest layers of the atmosphere.

A goal of our analysis is to move beyond the simple energy-balance view expressed in Eqn. 2 – which was previously applied to stratospheric aerosol forcing in McGraw & Polvani (2024) – because this approach typically treats precipitation responses as an automatic consequence of radiative perturbations without resolving the processes through which the atmospheric column restores energetic balance. To help narrow this gap, we here examine vertically-resolved latent heating profiles, equal to the vertical divergence of precipitation flux at each model level following conversion into energetic units via the scaling 28.9 W m$^{-2}$ per mm day$^{-1}$. By construction, vertically integrating this profile recovers the LΔP term in Eq. 2, allowing us to interpret that term as an aggregate response of processes operating across different levels of the troposphere.

## 3. Surface-Temperature-Independent Effects on Tropical Mean Precipitation

*a) The Stratospheric Perturbation*

We begin by identifying the mechanisms driving tropical precipitation responses to stratospheric sulfate in our simulations. Figure 2 summarizes the primary process interactions we examine: stratospheric sulfate perturbs radiative fluxes in the troposphere, altering the radiative heating/cooling structure, modulating convective activity, precipitation, and cloudiness, which in turn alters the radiative perturbation. Our analysis here uses vertically-resolved model output (Section 2c), offering previously unexplored insights into the processes involved. Radiative diagnostics are shown in Fig. 3, while responses in other relevant variables we will refer to are presented in Fig. 4. In this section, we focus on the GISS and mock-Walker SAM simulations, which capture large-scale tropical circulation and are therefore expected to provide the most reliable representation of SAI responses. Small-domain SAM simulations, which lack these large-scale dynamics, will be analyzed separately in Section 4.

As stratospheric aerosols influence tropospheric climate primarily through their radiative effects – the sole influence represented in our simulations – we begin by examining the aerosol radiative forcing itself. Figure 3 shows components of the aerosol radiative heating anomaly (base SAI minus control), with black lines representing the GISS simulations averaged over the tropics, teal lines for the same case in our SAM mock-Walker simulations, and grey lines for a GISS SAI simulation reran with only aerosol LW effects active. Starting with the aerosol radiative forcing (Fig. 3a), both models clearly show that stratospheric sulfate induces substantial heating of the stratosphere and upper troposphere, which are separated by cold-point tropopauses near 16 km. This stratospheric heating is attributable to absorption of outgoing longwave (LW) radiation by the sulfate layer, as is evidenced by similar stratospheric anomalies in the LW only case and the full base SAI case.

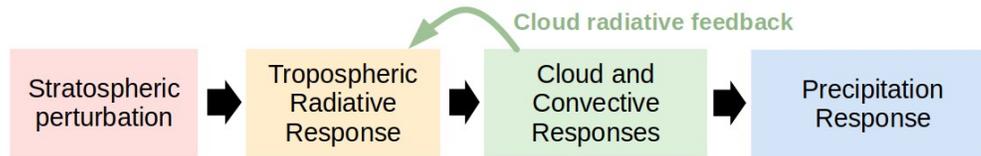

FIG. 2. Mechanisms driving surface-temperature-independent cloud and precipitation responses to stratospheric aerosols that are represented in our simulations and analyzed herein.

In Fig. 3b, we show the clear-sky radiative adjustment to the aerosol layer, which primarily reflects the impact of stratospheric warming as the aerosols absorb outgoing longwave (LW) radiation. Essentially, the aerosols trap LW energy in the stratosphere, but the resulting temperature increase (Fig. 4a) enhances radiation to space via the Planck response, offsetting much of the initial forcing. Consequently, the aerosol forcing (Fig. 3a) and clear-sky adjustment (Fig. 3b) largely cancel in the stratosphere, establishing a new equilibrium in the stratospheric energy budget.

*b) The tropospheric radiative response*

While the aerosol forcing and clear-sky adjustments largely cancel in the stratosphere, both produce net radiative heating in the troposphere. This heating arises from increased downward emission from a stratosphere that (1) contains an enhanced mass of longwave-emitting aerosols and (2) is anomalously warm and thus more emissive overall. Comparing the GISS base SAI case (black lines in Figs. 3a and 3b) to the LW only case (grey lines) in the troposphere shows that this heating is driven by LW effects. Clear-sky tropospheric adjustments (Fig. 3b) also reflect water vapor changes, though this is likely a secondary influence compared to that of the large stratospheric temperature anomaly.

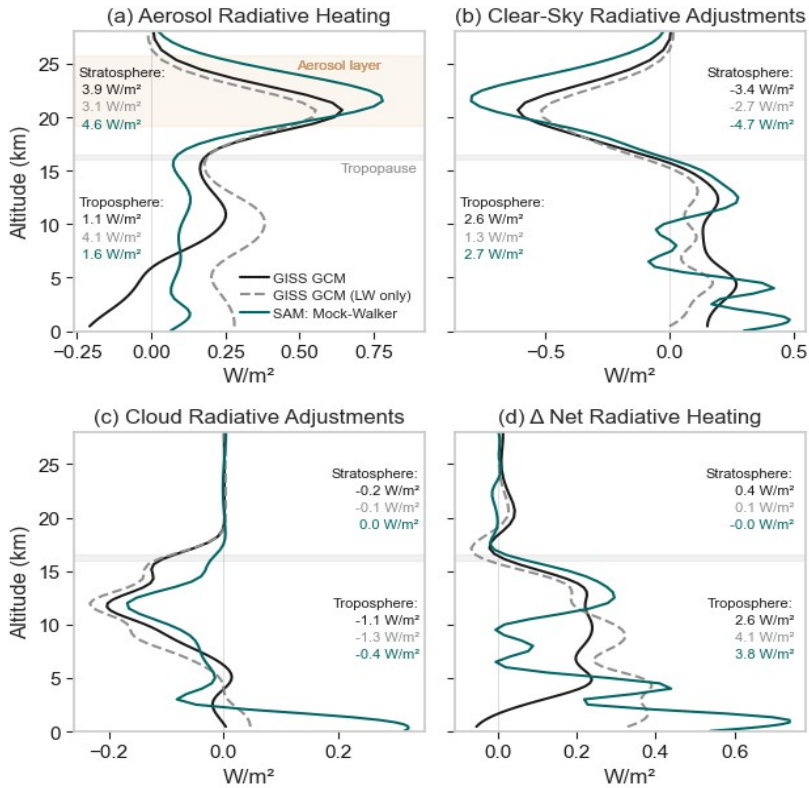

FIG. 3. Vertically-resolved atmospheric radiative flux responses to SAI in the GISS and SAM mock-Walker simulations. Results are smoothed with a 1 km Gaussian filter to improve readability by reducing sharp features.

The LW only case produces more tropospheric heating than the base SAI case, indicating that aerosol shortwave (SW) effects oppose this warming. As noted previously (Laakso et al., 2024; McGraw & Polvani, 2024), this SW cooling arises because stratospheric aerosols reflect sunlight that would otherwise be absorbed by the underlying atmosphere. We conclude that the primary SW mechanism is reduced absorption by water vapor, which is the atmosphere's dominant absorber at these wavelengths (Kiehl & Trenberth, 1997). Supporting this, elevated aerosol heating when SW effects are omitted is also robust in our SAM simulations (as shown in Section 7), which include no tropospheric aerosols, and is evident in clear-sky aerosol heating profiles (not shown). Nevertheless, LW effects dominate overall, as indicated by the positive aerosol radiative heating in both models (Fig. 3a).

A second factor contributing to reduced tropospheric heating is the cloud response to the overlying aerosol layer. We will examine the physical mechanisms that lead to this cloud response below, and here simply emphasize the reduced heating attributed to cloud radiative adjustment, shown in Fig. 3c. The three radiative terms – direct aerosol forcing, clear-sky adjustment, and cloud adjustment (Figs. 3a-

c) – sum to the total aerosol-induced radiative response (Fig. 4d), which in the troposphere is positive in both models. In summary, the warmer, aerosol-rich stratosphere emits more downward radiation, producing tropospheric heating, which is partially offset by reduced solar absorption from water vapor and diminished cloud radiative heating.

*c) The response of clouds and convection*

We now examine the responses of convection and cloudiness to the above-described radiative responses caused by the presence of stratospheric aerosols. The aerosol-induced net heating in the upper troposphere (Fig. 3b) stabilizes the atmospheric column, weakening convective updrafts and reducing the vertical advection of air mass at these altitudes, as is shown in Fig. 4b. This diminished convective mass flux overall limits the transport of moisture and energy from the lower to upper troposphere.

Consistent with this weaker convection, upper-tropospheric cloud amount is reduced (Fig. 4c). In addition to the convective driver, cloud amount is likely influenced by the warming itself (Fig. 4a): higher temperatures increase the saturation specific humidity, requiring moister air parcels to reach condensation. As a result, even updrafts that still reach high altitudes produce less condensation, further reducing cloud formation. These combined effects represent a surface-temperature-independent adjustment of convection and upper-tropospheric clouds in response to stratospheric aerosols. Cloud responses will be discussed in more detail in Section 4.

*d) The mean precipitation response*

SST-independent SAI effects are found to reduce mean precipitation by 2.1% in our SAM mock-Walker base SAI experiment and by the same percentage in our GISS base SAI experiment averaged across the tropics. We show the latent heating profiles in Fig. 4d. Column-integrated values are labeled in both latent energy and precipitation units, with this integral equaling the total amount of precipitation reaching the surface (see Section 2d for more explanation).

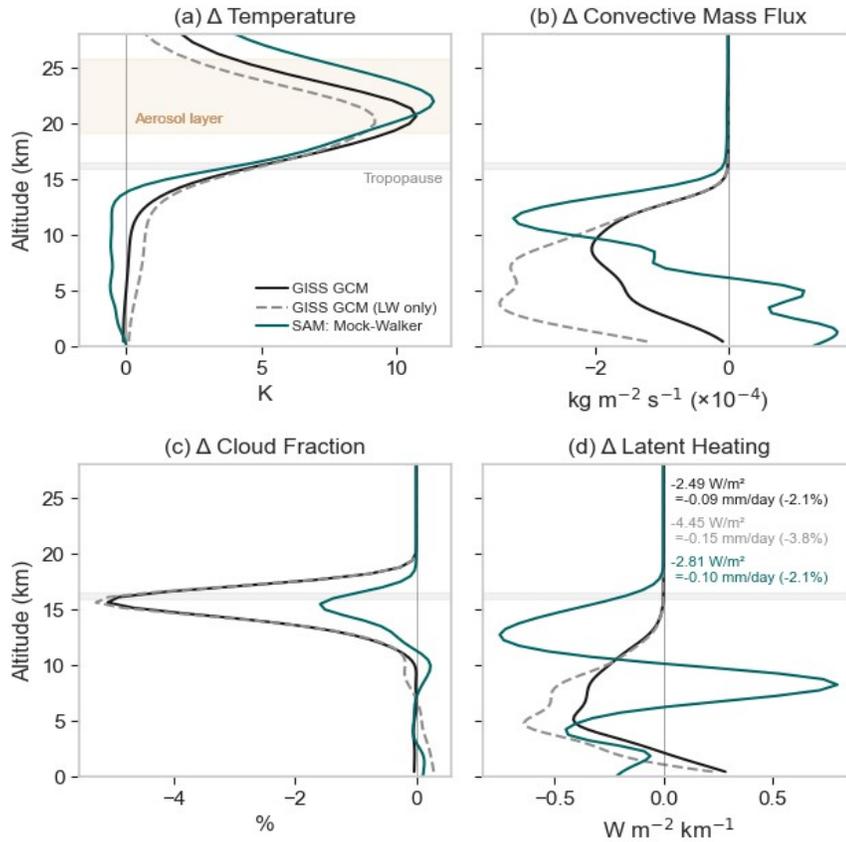

**FIG. 4.** Key atmospheric response variables in the GISS and SAM mock-Walker simulations, analogous to Fig. 3. Results are smoothed with a 1 km Gaussian filter.

The reduction in latent heating (Fig. 4d) largely offsets the increase in radiative heating of the troposphere (Fig. 3d). To better understand how the models arrive at this latent heating reduction, it is useful to consider the intermediate processes that reconcile radiative and heating responses. In GISS (black lines), the cancellation is relatively straightforward: SAI induces radiative heating at nearly all vertical levels (see Fig. 3d) which weakens convection throughout the troposphere (Fig. 4b) by suppressing atmospheric buoyancy, and thereby reduces vertical moisture transport and condensation rates.

In SAM, reductions in latent heating also generally offset tropospheric radiative heating, though the vertical structure is more complex than in GISS. From ~10 km to the tropopause latent heating decreases, yet the mid-troposphere experiences intensified latent heating despite minimal radiative heating response to SAI here (Fig. 4d). The latent heating increase may reflect enhanced convective transport of moisture from the lower atmosphere (Fig. 4b) or reduced evaporative cooling due to less precipitation falling from above. In the lower troposphere, radiative heating strengthens local convection, with reduced latent heating reflecting moisture loss to layers aloft.

It remains unclear which model more accurately represents SAI impacts on the tropical water cycle. Although both simulate a 2.1% reduction in tropical-mean precipitation, this agreement masks substantial differences in how radiative and latent heating compensate through the troposphere. In the GISS GCM, parameterized convection effectively converts radiative heating into reduced precipitation. SAM produces a more intricate vertical response driven by convection-resolving dynamics, though it is unclear whether this complexity is more realistic or simply different. We also have not discussed boundary-layer drizzle processes, which further complicate interpretation: these are sensitive to radiative heating because cloud-top radiative cooling drives circulations linking low clouds to surface moisture (Bretherton & Wyant, 1997), yet even the 3 km resolution of our SAM simulations is insufficient to reliably simulate these processes. Despite these uncertainties, this comparison clarifies the processes that reconcile radiative and latent heating and highlights the aspects of model physics that must be captured reliably to accurately predict SAI impacts.

**4. Cloud Radiative Heating and Insights for Constraining Precipitation Responses**

*a) Modeling challenges for projections of SAI hydroclimate impacts*

Cloud radiative effects are a well-known source of uncertainty in climate projections and likely drive much of the intermodel disagreement in Earth's mean hydrological response to surface warming (McGraw et al., 2025). We here demonstrate that these effects also present a major obstacle to accurately assessing the hydroclimate risks of SAI. As shown in Section 3, simulated upper-tropospheric cloud amount declines in response to stratospheric aerosols, and the resulting reduction in cloud radiative heating partially offsets the initial tropospheric energy response to the aerosols. In this section, we examine differences in the cloud radiative heating response across the different models used in this study. Here we focus on the various SAM configurations to isolate how differences in cloud properties within a single model affect the response, using GISS only as a reference point for comparison.

We will firstly show key differences in cloud properties across our various model configurations, and then will discuss (in Section 4b) how these relate to distinct responses to SAI. To recap our simulation setups, the GISS global model uses a coarse grid with parameterized convection, whereas the SAM simulations resolve convection in limited-area 3D domains. The SAM configurations include:

(a) a mock-Walker configuration at 3 km resolution with a 5°C SST gradient imposed along a 6144 km axis, (b) a small-domain configuration with uniform SSTs at 3 km resolution, and (c) a small-domain configuration as in (b) but at 1 km resolution. While all configurations might be considered suitable for use in SAI risk evaluations, we highlight their limitations as a cautionary note on potential overconfidence in claiming definitive results.

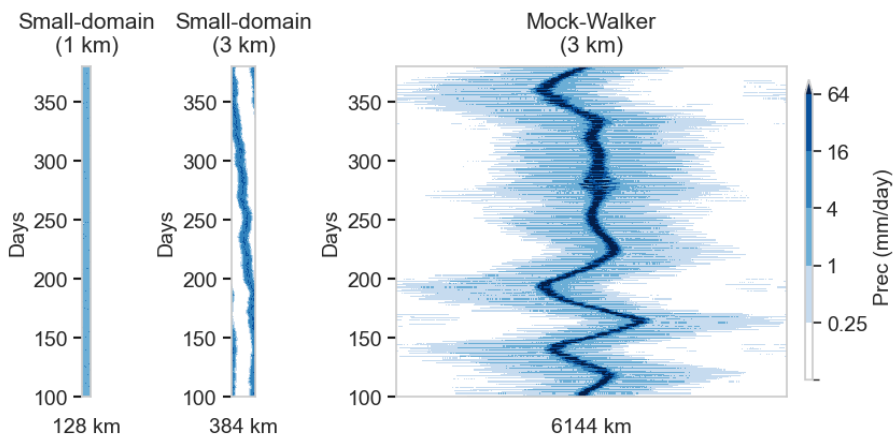

FIG. 5. Spatiotemporal structure of precipitation in each SAM configuration used herein. Shown are Hovmöller diagrams of mean precipitation across the horizontal x-axis from days 100-380 in the control (no aerosol) simulations. For the mock-Walker simulation, this corresponds to the longer of the two horizontal axes and SST is hottest in the center of the x-axis presented in the image.

While limited-domain RCE models have become widely adopted in climate studies as a more elegant alternative to coarse-grid GCMs, it is important to note that such models are idealized, meant to explore simplified scenarios rather than a close representation of tropical climate. This is evident in the horizontal structure of precipitation in our SAM control simulations (Fig. 5). The small-domain simulation at 1 km resolution produces nearly uniform precipitation whereas, despite uniform SSTs, at 3 km resolution the same model shows precipitation concentrated into a band due to convection self-organizing (as in evident in Fig. 1a, with this process explained in Wing et al., 2018b). In the mock-Walker simulations, the imposed SST gradient drives a large-scale circulation that produces a yet narrower ascent region over the warm center of the domain. However, the location of maximum ascent shifts chaotically relative to the SST pattern, reproducing a previously noted structure (Dagan et al., 2023; Wing et al., 2024) that appears to reflect a combination of forced and self-organizing convection. Although the extent to which these organization patterns reflect real-world processes rather than model artifacts remains uncertain, this organization exerts a strong influence on cloud properties that, as we will show, substantially affect the simulated response to SAI.

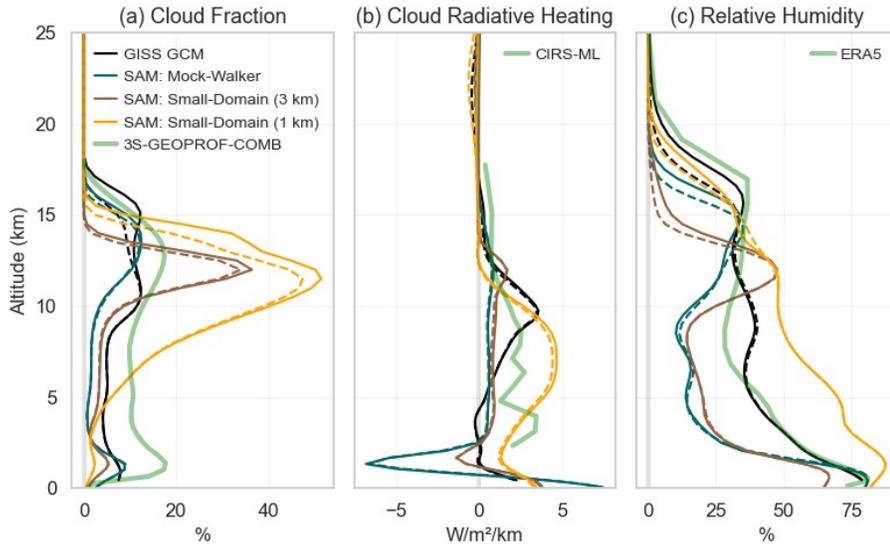

FIG. 6. Cloud fraction, radiative heating, and relative humidity in our simulations. Solid lines are profiles in our control simulations, and dashed lines of the same color are from the matching base SAI experiment, with the exception that green lines depict observational products for comparison. All relative humidity values are relative to liquid water. All values in (a) and (c) are averages across the tropics from 20°S to 20°S, consistent with the rest of this study, while in (b) all averaging is done from 15°S to 15°S to be consistent with the CIRS-ML observational product as processed in Stubenrauch et al. (2021) and previously compared to SAM simulations in Gasparini et al. (2024).

In Fig. 6, we show vertical profiles of cloud fraction, cloud radiative heating, and relative humidity in our control simulations (solid lines, excluding green), which reveal large intermodel differences and significant biases relative to observational benchmarks (green lines), themselves subject to uncertainty. These observational products encompass cloud fraction from the combined radar-lidar 3S-GEOPROF-COMB product (Bertrand et al., 2024), cloud radiative heating from CIRS-ML (Stubenrauch et al., 2021), and relative humidity (RH) from ERA5 reanalysis (Hersbach et al., 2020). The small-domain SAM simulations exhibit the strongest biases: the 1 km configuration (orange lines) exaggerates cloud fraction and radiative heating in the mid- to upper troposphere, while all SAM configurations are found to underestimate cloud occurrence in the lower troposphere. The 3 km small-domain case (brown) shows the opposite (negative) bias in cloud radiative heating, likely because convection remains confined to a narrow part of the domain (Fig. 5). Similar deficiencies are seen in the mock-Walker configuration (blue), which additionally features substantial radiative cooling from developed low cloud layer in subsidence zones. This low cloud radiative cooling absent in the other setups and uncertain in observations, while global models indicate an uncertain mix of cooling and heating low cloud layers (Voigt et al., 2024) and kernels factor low clouds as overall radiatively cooling (Zhang et al., 2021). Further affecting hydrological cycle representation, relative humidity biases differ sharply across configurations. The unaggregated (1 km small-domain) SAM setup produces strong negative RH biases, while the aggregated (3 km small-domain and mock-Walker) setups show strong positive biases. For reference we also include the GISS model here. Despite its coarse grid and parameterized convection, GISS aligns best to the three observational products, though no model is very close and the observational uncertainties are substantial. These comparisons highlight that, although convection-resolving models offer valuable mechanistic insights, reliably producing a representative tropical atmospheric structure remains a substantial challenge.

*b) Dependence of mean tropical precipitation response on cloud representation*

In Figs. 7 and 8, we show radiative flux responses to SAI (base SAI experiments minus control) and key response variables in each of the three SAM configurations. Although these simulations use the same model under an identical aerosol perturbation, the resulting atmospheric radiative anomalies differ substantially. Variations in the aerosol forcing (Fig. 7a) and clear-sky adjustment (Fig. 7b) likely arise from differences in the emission temperature of upward longwave radiation to the aerosol-rich stratosphere, given sensitivity to the differences in high cloud amount. Interestingly, despite lying at opposite ends of the model complexity spectrum, the 1 km small-domain simulations (orange lines) and the GISS global model (black lines in Fig. 3) produce broadly similar aerosol and clear-sky responses. Yet the 1 km configuration is our only configuration that exhibits a negative total radiative heating response (Fig. 7d). This reflects a large disagreement in the cloud-radiative heating response, which more than cancels the aerosol forcing in the 1 km case (gold line in Fig. 7c) but has almost no column-integrated effect in our SAM mock-Walker configuration (teal line) and is also weak in GISS (black line in Fig. 3c).

The different cloud responses produce precipitation anomalies (Fig. 8d) that vary not only in magnitude but even in sign across configurations. Notably, the 1 km small-domain configuration is the only case in which tropical-mean precipitation increases under SAI. However, this configuration also exhibits a substantial positive bias in base-state cloud radiative heating (Fig. 6b), which likely inflates its precipitation response; the reductions simulated in the other configurations are therefore more credible. Across simulations, cloud radiative heating responses (Fig. 7c) scale similarly to both their base-state values (Fig. 6b) and the corresponding cloud-fraction responses and base states (Figs. 8c and 6a). Overall, these results show that simulated hydroclimate responses to SAI are highly sensitive to cloud behavior, highlighting the need for accurate cloud representations when assessing SAI risks.

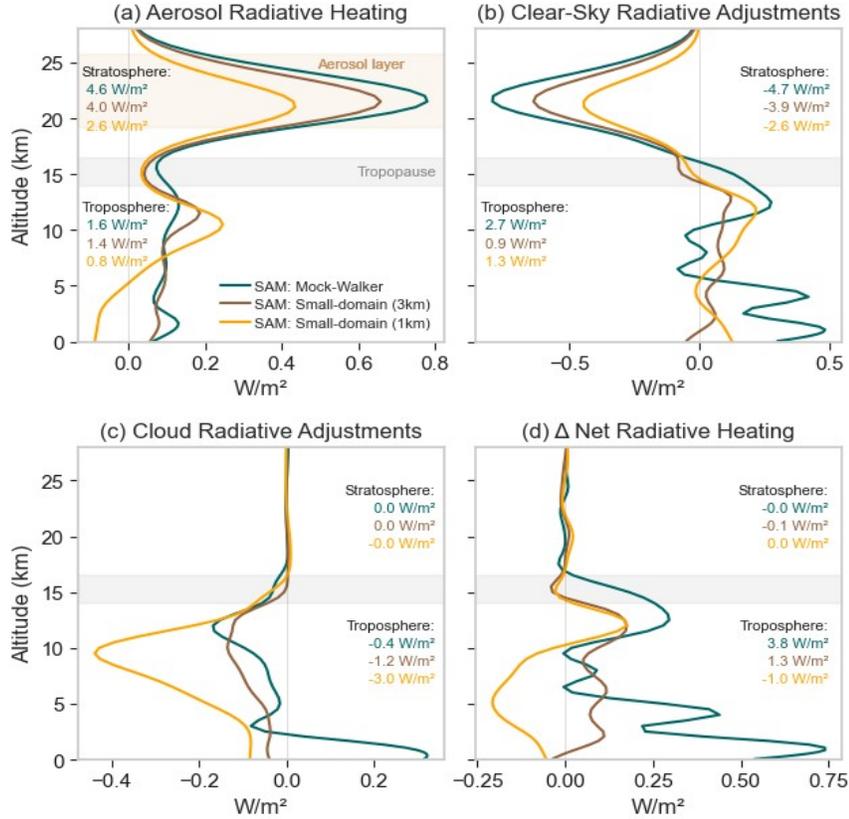

FIG. 7. As in Fig. 3 but showing radiative flux responses across the three SAM configurations.

**5. Lack of a Tropical Circulation Response from Surface-Temperature-Independent Effects**

Ferraro et al. (2014) argued that stratospheric aerosol injection weakens the tropical overturning circulation through surface-temperature-independent SAI effects reflecting increased atmospheric stability. In their analysis, this was expressed as a narrowing of the distribution of 500 hPa vertical velocity ($\omega 500$), reflecting reduced contrast between regions of strong ascent and descent when SAI was applied. Because $\omega 500$ captures the strength of mid-tropospheric vertical motion, its distribution provides a compact measure of circulation intensity. Here we explore the response of the tropical circulation to SAI using our global and convection-resolving mock-Walker simulations and revisit the suggestion of Ferraro et al. (2014).

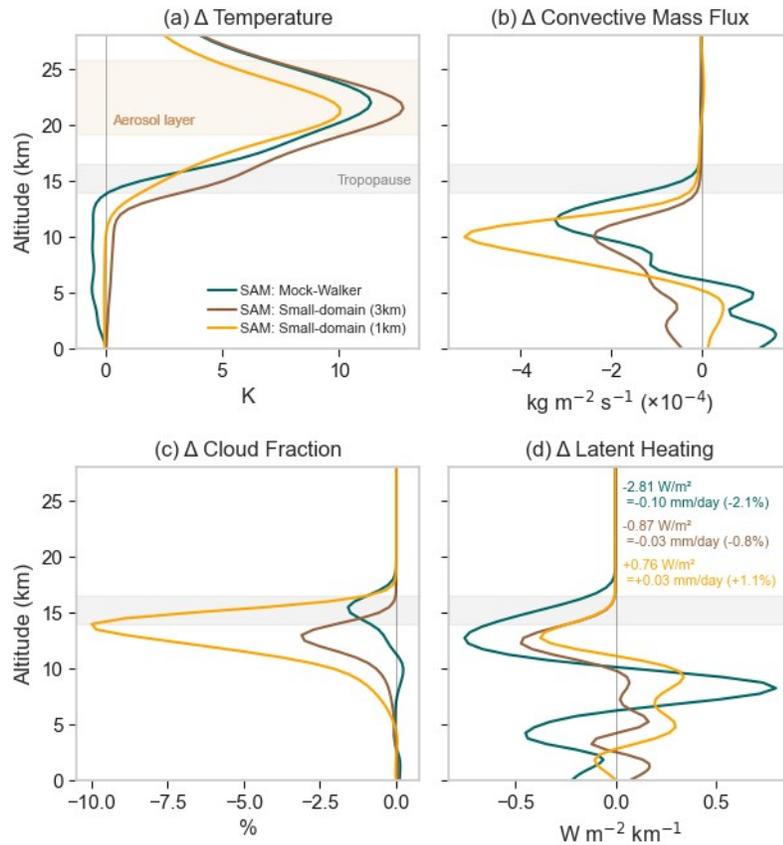

FIG. 8. As in Fig. 4 but showing key atmospheric response variables across the three SAM configurations.

To quantify circulation changes, we follow Ferraro et al. (2014) in sorting simulated regions into deciles based on their mean control ω500, and then we evaluate how SAI modifies the distribution. In GISS, we focus on mean regional responses to SAI by ranking ω500 across tropical model columns after 12-year averaging, then comparing the base simulation to the control at each location. For the SAM mock-Walker experiments, we use daily output to track the migrating ascent region (Fig. 5) in both the control and base SAI simulations. For each of the 150 assessed days in each simulation, ω500 deciles are calculated, and then for all days the matching deciles are averaging together. This approach allows us to assess circulation anomalies without confounding effects from the ascent region shifting to different locations in the base SAI simulation as in the control run.

Our results (Fig. 9, orange lines) show no detectable narrowing of the ω500 distribution in the SAM mock-Walker experiments, and only a weak reduction in the GISS simulations. Moreover, comparing control ω500 and the SAI-induced change Δω500 across model columns (not shown) indicates at most a weak systematic relationship, with a coefficient of determination ($R^2$) of merely 0.19. In the following section, we will show that even the small response in our GISS simulations likely lacks a genuine surface-temperature-independent mechanism, as it arises from aerosols being able to cool over land but not oceans in our prescribed-SST simulations, producing exaggerated horizontal temperature gradients.

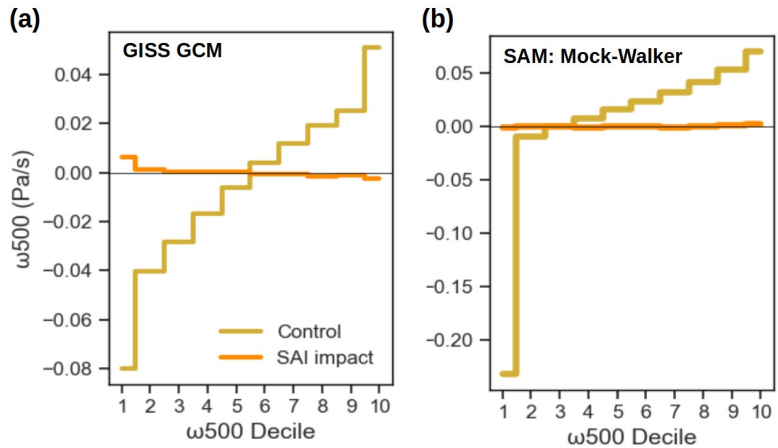

FIG. 9. Mid-tropospheric vertical motion (ω500) in GISS and SAM mock-Walker simulations. Gold lines show the decile-ranked ω500 distribution in the control simulations, and orange lines show the change induced by the base SAI experiment relative to the control. This approach, which follows Fig. 4 of Ferraro et al. (2014), highlights any potential redistribution of circulation.

To reconcile our findings with Ferraro et al. (2014), it is worth noting that their circulation responses were simulated with a temperature-responsive slab ocean, permitting the circulation to respond to surface temperature gradients produced by the aerosols. These gradients may have been considerable given that their aerosol distribution (their Fig. 1a) was concentrated near the equator, and hence the expected cooling would offset the temperature gradient that drives the overturning circulation. In contrast, our SAM simulations isolate purely radiative aerosol effects, and our GISS simulations only include a surface cooling response over land. Furthermore, in our simulations, which test results in two separate radiation schemes (GISS and RRTM), SAI does not produce warming throughout the troposphere (see Figs. 4a and 8a) as in the intermediate-complexity model of Ferraro et al. (2014), limiting conditions under which a surface-temperature-independent circulation response could emerge. Overall, our results indicate that when surface-temperature-independent effects are fully isolated, no clear tropical circulation response is seen.

## 6. Regional Precipitation Responses from Surface-Temperature-Independent Effects

We next examine regional precipitation responses to SAI in the GISS and SAM mock-Walker simulations, going beyond our tropical-mean analysis in Section 3d. In addition to identifying how precipitation changes vary across the tropics, we here investigate the mechanisms that enable regional anomalies to exceed the tropical-mean response.

We first separate regions of ascent and descent, by applying the $\omega 500$-binning framework introduced above. In GISS, ascent and descent regions are defined by splitting the control-simulation distribution of $\omega 500$ into equal-area halves (see Fig. 9). In the SAM mock-Walker simulations, ascent is instead identified as the 20% of grid points having the most negative $\omega 500$ values and descent as the remaining 80%; binning is then applied to track horizontal shifts in the $\omega 500$ distribution over time (Section 5). Figure 10 shows the resulting column-integrated energetic responses to SAI, with red, blue, and green bars representing radiative, latent, and sensible heating contributions (the terms of Eqn. 2, which are all standard model outputs). Regional differences in precipitation are reflected in the latent heating response (blue bars, noting from Section 2c that 29.8 W m$^{-2}$ integrated latent heating = 1 mm day$^{-1}$ surface precipitation).

Both models display substantial regional variations in the precipitation response to SAI. Precipitation declines most strongly in the ascent regions, consistent with these areas hosting the deepest clouds and strongest convection. Yet the strength of these reductions is somewhat counterintuitive, especially in the SAM mock-Walker configuration (bottom row). Here the latent heating reduction in ascent regions is far stronger than the local radiative heating enhancement, such that ascent regions are overall losing energy through diabatic processes (compare leftmost blue, red, and grey bars). Further, in SAM there is no clear circulation adjustment (see Fig. 9a) that could explain how energy is redistributed across the domain. While the dominant mechanism remains uncertain, we propose three potential explanations for how energetically imbalanced responses to SAI might persist without a circulation response: (1) the fixed circulation continually replaces cooling (heating) air in ascent (descent) regions with air that has experienced the opposite energetic tendency, preserving the imbalances; (2) horizontal migration of the ascent region (Fig. 5) could shift local energetic anomalies across the domain; and (3) enhanced precipitation in descent zones (middle blue bar on the lower row of Fig. 10) reduces the availability of water advected to ascent region, enhancing the latent heat reduction in the ascent region – a mechanism reminiscent of but opposite to the aerosol response described by Dagan et al. (2023), which we note did involve altered circulation.

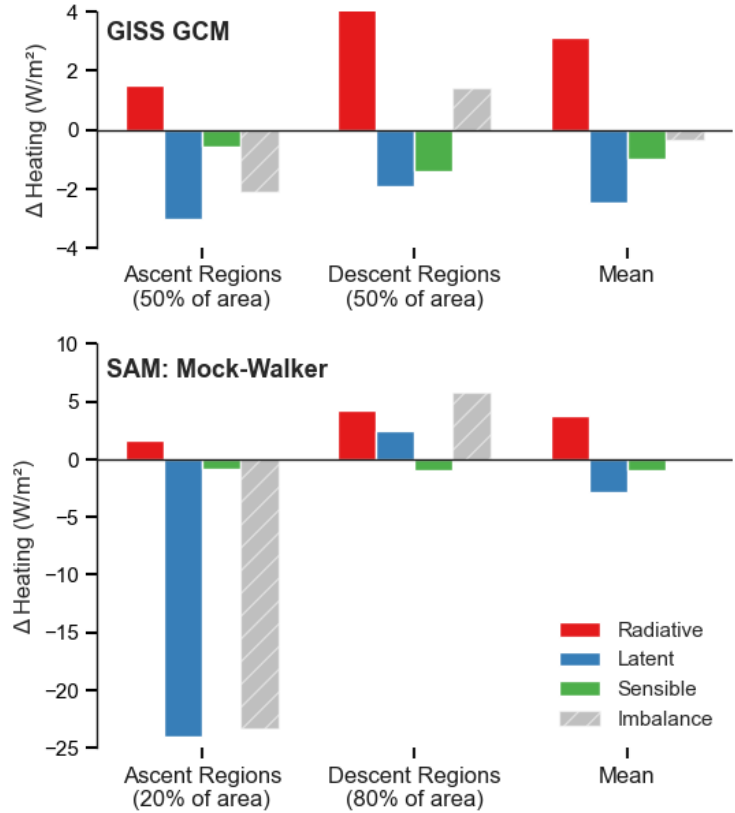

FIG. 10. Energetic responses to SAI separated into ascent and descent regions. Bars show radiative (red), latent (blue), and sensible (green) contributions; grey hatching denotes the sum of all components. Ascent and descent regions in SAM are defined in a Lagrangian sense, following shifts in the ω500 distribution.

We next consider the GISS simulations. Although differences in the precipitation response between ascent and descent regions are smaller here than in SAM (compare blue bars across the rows of Fig. 10), a more detailed analysis reveals distinct spatial differences. Figure 11 presents maps of the energetic responses to SAI, expressed in precipitation units following the energetic framework of Muller and O'Gorman (2011, their Eqn. 3). This method estimates the contribution of dynamical advection to precipitation anomalies based on the response of ω along with the dry static energy field of the control climate. While the radiative response (Fig. 11a) is seen to be the most consistent energetic response across regions, the spatial pattern of precipitation anomalies (Fig. 11e) aligns most closely with dynamical contributions (Fig. 11c). However, there is reason to view these dynamically-modulated responses cautiously: they display a tendency toward ascent over tropical continents and subsidence over most ocean regions, and hence are likely caused by these prescribed-SST simulations still cooling over land in response to the aerosols, resulting in temperature gradients that drive land-oceans circulations. We note that while it might be advantageous to prescribe land temperatures as we do SSTs, this is technically challenging to accomplish in global models (Andrews et al., 2021).

Dynamically-induced precipitation responses like in Fig. 11c may be a physically plausible response to stratospheric aerosols from large volcanic eruptions, wherein aerosols form rapidly, allowing pronounced land–ocean gradients to develop (c.f., McGraw and Polvani, 2024). However, in the case of gradual, sustained SAI deployment, they are unlikely. Instead, these results highlight a specific risk pathway: unsuitably rapid aerosol injections could cool land while outpacing the ability for SSTs to catch up, producing temporary land-ocean temperature gradients and hydroclimate anomalies.

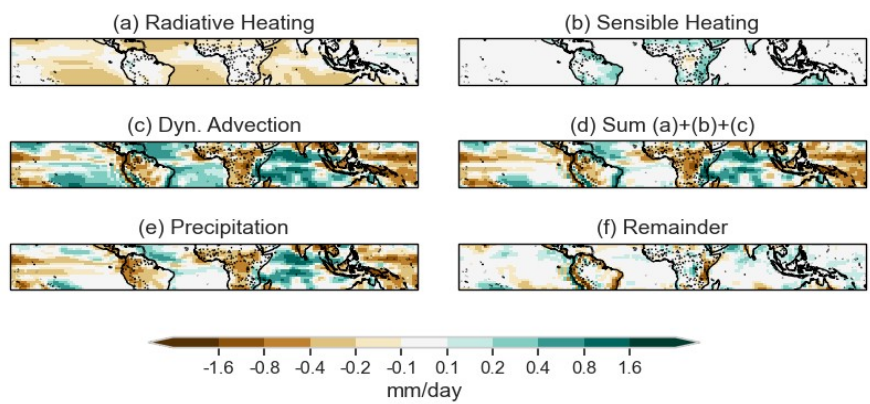

FIG. 11. Energetic responses to SAI in the GISS model base SAI experiment, in precipitation units and × -1 to express estimated influences of each term on the simulated precipitation change. Shown are the theoretical influences of radiative heating (a), sensible heating (b), and dynamical advection (c, following the method of Muller and O'Gorman, 2011) on mean regional precipitation, along with their sum (d), actual simulated precipitation change (e), and the residual calculated as (e)-(d).

## 7. Sensitivity of Mean Tropical Precipitation Response to Aerosol Particle Size and Altitude

Having established the responses in our SAI base experiment, we now examine how these differ when the aerosol layer is placed closer to the tropopause or composed of smaller particles. These sensitivity experiments provide insight into whether SAI impacts can be deliberately optimized – or "steered" – and into the predictability of outcomes under specific implementation strategies. Fig. 12 shows column-integrated radiative, sensible, and latent heat flux responses from these experiments, alongside the longwave only case examined in Section 3 for reference. Radiative responses are also shown over the troposphere, as overlying hollow hatched bars.

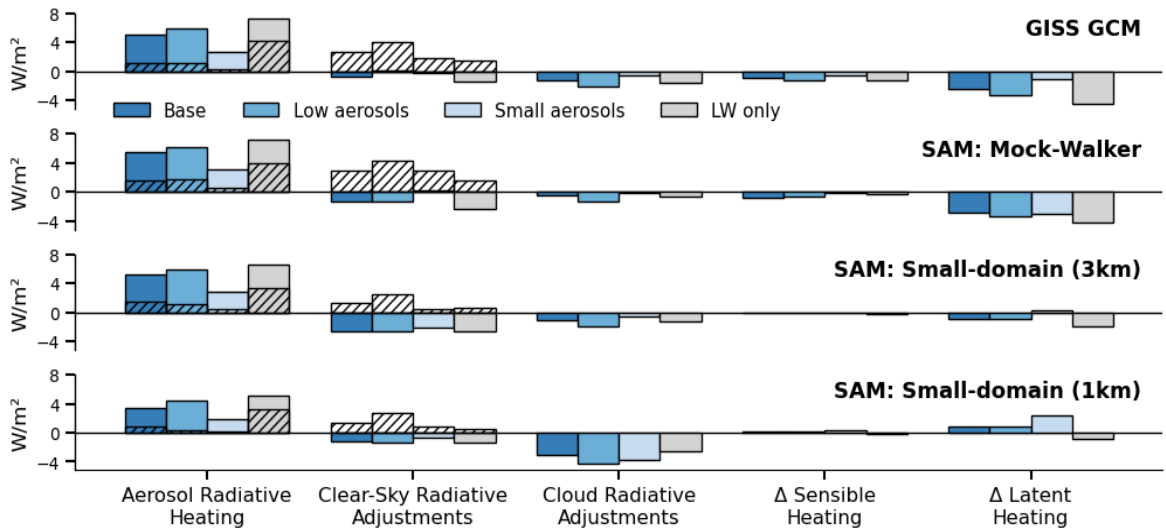

FIG. 12. Column-integrated (filled bars) and tropospheric (hatched bars) energetic responses from the base SAI experiment, sensitivity experiments varying aerosol altitude and particle size, and the LW only case. Minor differences from the values in Figs. 3, 4, 7, and 8 reflect corrections to ensure consistency with simulated top-of-atmosphere and surface fluxes, whereas those earlier values were based wholly on manual vertical integration of the profiles.

We first consider the effect of lowering the aerosol layer toward the tropopause (see Section 2c for details). All model versions show that this scenario produces slightly stronger aerosol radiative heating across the atmospheric column than our base SAI case (compare dark and medium blue bars in the leftmost column), but with a less distinct difference in the tropospheric aerosol heating (overlying hatched bars). Concentrating the aerosol layer nearer the tropopause causes stronger lower-stratospheric warming, which more clearly enhances tropospheric heating (see hatched bars in the second group). However, this stronger radiative perturbation also intensifies the SAI-induced reduction in cloud amount (not shown), with the resulting cloud radiative response (third group of bars) negating much of the radiative heating difference. The two models that most thoroughly represent tropical processes – GISS and SAM mock-Walker (top two rows) – both show an enhanced precipitation decline in the low aerosols case, as reflected by the latent heating responses (rightmost column; noting as explained in Section 2c that 1 mm/day of precipitation corresponds to 29.8 W m$^{-2}$ of latent heating).

This is consistent with the altitude sensitivity identified in CESM simulations by Usha et al. (2024). By contrast, the absence of a matching response in the small-domain SAM simulations (bottom two rows) highlights the difficulty in constraining the hydrological influence of aerosol height.

We next examine the case with smaller aerosol particles (see Section 2c for details). Because smaller sulfate particles tend to scatter more sunlight (Lacis, 2015), they more strongly inhibit shortwave absorption by tropospheric water vapor, weakening the aerosol radiative heating relative to the SAI base case (light versus dark blue in the leftmost column). This behavior contrasts with the LW only experiment (grey bar), wherein removing aerosol shortwave effects strengthens tropospheric heating. Likewise, the small particle scenario produces weaker precipitation reductions in most models, as reflected by the weaker latent heating reduction in the rightmost column of Fig. 12. However, the SAM mock-Walker shows little sensitivity of mean precipitation to particle size, reflecting offsetting cloud-radiative and sensible heating responses. Taken together, these results suggest that an elevated aerosol layer composed of small particles would optimally minimize tropical-mean precipitation reductions from surface-temperature-independent effects. Yet this conclusion should be viewed cautiously, as it hinges on delicate balances between poorly constrained atmospheric processes.

## 8. Discussion and Conclusions

Our analyses provide a more controlled and in-depth evaluation of surface-temperature-independent atmospheric responses to stratospheric aerosols than has previously been available. Across both convection-resolving and global simulations, we pinpoint consistent atmospheric responses: an aerosol-rich, warmed stratosphere radiates extra heat into the troposphere, overall suppressing convection and cloudiness – which in turn reduces the tropospheric radiative heating excess – and through these changes alters precipitation. The outcome is an overall reduction in tropical mean rainfall accompanied by sharper regional anomalies, yet will prove difficult to constrain due to dependence on the uncertain cloud radiative heating response. Contrary to earlier suggestions, we do not find robust surface-temperature-independent changes in the strength of the tropical overturning circulation.

These results provide several key insights into SAI risks. First, they corroborate earlier findings of reduced tropical precipitation from surface-temperature-independent effects (Ferraro et al., 2014; Usha et al., 2024), here extended to high-resolution simulations that explicitly resolve convection rather than relying solely on parameterizations. Second, our results highlight the difficulty of reliably predicting SAI impacts, as cloud radiative processes are both strong and poorly constrained – so much so that in one model configuration (i.e. 1km-resolution small-domain SAM) they were sufficient to reverse the sign of the precipitation response. Sensitivity tests indicate that SAI hydroclimate outcomes could, in principle, be moderated by adjusting aerosol altitude or particle size, yet current models likely lack the fidelity to ensure that such optimization would be effective. This raises the concern that any SAI deployment might proceed without sufficient knowledge of its hydroclimate consequences. Further, substantial regional precipitation responses have been identified, underscoring the potential for disruptive local hydroclimate impacts even in the absence of circulation changes. Finally, despite the lack of a true surface-temperature-independent circulation response, our prescribed-SST simulations suggest that a 'fast' circulation response could emerge if SAI were deployed rapidly enough that land cooled before sea surface temperatures could respond, temporarily altering land-ocean circulations.

These findings also inform our understanding of volcanic aerosol impacts, which involve the same surface-temperature-independent atmospheric response mechanisms. Convection-resolving simulations of volcanic aerosol layers would be nearly identical to the stratospheric sulfate experiments presented here for which, we note, the aerosol optical depth is comparable to what followed the 1815 eruption of Mt. Tambora. Our results support the conclusion that volcanic aerosols exert a genuine surface-temperature-independent precipitation reduction, consistent with McGraw & Polvani (2024), who reached similar findings using prescribed-SST simulations with a radiative-kernel analysis. Notably, we found in the present study that GCM simulations with only SSTs prescribed – allowing land temperatures to adjust freely – can produce the same magnitude of drying as an ocean-only fixed-SST model. What remains less understood is how these hydroclimate responses evolve in time. Unlike controlled SAI forcing, volcanic forcing is intrinsically transient: sulfur gases convert to aerosols and are subsequently transported poleward through the stratosphere. As noted in Section 6, regional precipitation and circulation responses like those seen in our prescribed-SST global simulations may be dominant effects during the early months following a large eruption.

Our findings highlight continued challenges that must be addressed to constrain SAI climate impacts, including the need for a rigorous reduction of uncertainties stemming from model representation of clouds. Here we have taken steps in this direction by using cutting-edge simulations that resolve convection to quantify process sensitivities across both the vertical and horizontal structure of the atmosphere. Future progress could similarly combine convection-resolving simulations, which isolate physical mechanisms, with global simulations that capture broader Earth system feedbacks – potentially addressing both perspectives with a single observationally-constrained global storm-resolving model. Only by narrowing uncertainties in cloud radiative processes – increasingly recognized as a key control on rain rates (Harrop & Hartmann, 2023; Haslehner et al., 2024; McGraw et al., 2025) – can we gain confidence in projecting how stratospheric aerosols influence the troposphere and alter precipitation patterns.

*Acknowledgments.*

The authors thank Andrew A. Lacis, Gregory S. Elsaesser, and Blaz Gasparini for helpful conversations, and Benjamin Goldman for helping test the vertically-resolved radiation diagnostics in GISS ModelE.

*Data Availability Statement.*

The SAM model is available for download at http://rossby.msrc.sunysb.edu/SAM/, while the GISS model code is available at https://simplex.giss.nasa.gov/snapshots/. The model output from GISS and SAM simulations used in this study has been made publicly available on a Zenodo archive at https://doi.org/10.5281/zenodo.17643252.

## REFERENCES

Andrews, T., P.M. Forster, O. Boucher, N. Bellouin, and A. Jones, 2010: Precipitation, radiative forcing and global temperature change. Geophys. Res.Lett., 37, L14701, https://doi.org/10.1029/2010GL043991

Andrews, T., C. J. Smith, G. Myhre, P. M. Forster, R. Chadwick, and D. Ackerley, 2021: Effective radiative forcing in a GCM with fixed surface temperatures. J. Geophys. Res. Atmos., 126, e2020JD033880, https://doi.org/10.1029/2020JD033880

Bala, G., K. Caldeira, and R. Nemani, 2010: Fast versus slow response in climate change: Implications for the global hydrological cycle. Climate Dynamics, 35(2), 423–434, https://doi.org/10.1007/s00382-009-0583-y

Bertrand, L., J. E. Kay, J. Haynes, and G. de Boer, 2024: A global gridded dataset for cloud vertical structure from combined CloudSat and CALIPSO observations. Earth Syst. Sci. Data, 16, 1301–1316, https://doi.org/10.5194/essd-16-1301-2024

Bretherton, C. S., & Wyant, M. C. (1997). Moisture transport, lower-tropospheric stability, and decoupling of cloud-topped boundary layers. Journal of the Atmospheric Sciences, 54(1), 148–167. https://doi.org/10.1175/1520-0469(1997)054<0148:MTLTSA>2.0.CO;2

Crutzen, P. J., 2006: Albedo enhancement by stratospheric sulfur injections: A contribution to resolve a policy dilemma? *Clim. Change*, 77, 211–219, https://doi.org/10.1007/s10584-006-9101-y

Ferraro, A. J., E. J. Highwood, and A. J. Charlton-Perez, 2014: Weakened tropical circulation and reduced precipitation in response to geoengineering. *Environ. Res. Lett.*, 9, 014001, https://doi.org/10.1088/1748-9326/9/1/014001

Ferraro, A. J., and H. G. Griffiths, 2016: Quantifying the temperature-independent effect of stratospheric aerosol geoengineering on global-mean precipitation in a multi-model ensemble. *Environ. Res. Lett.*, 11, 034012, https://doi.org/10.1088/1748-9326/11/3/034012

Gasparini, B., G. Mandorli, C. Stubenrauch, and A. Voigt, 2024: Basic physics predicts stronger high cloud radiative heating with warming. Geophys. Res. Lett., 51, e2024GL111228, https://doi.org/10.1029/2024GL111228

Hansen, J., A. Lacis, R. Ruedy, and M. Sato, 1992: Potential climate impact of Mount Pinatubo eruption. *Geophys. Res. Lett.*, 19, 215–218, https://doi.org/10.1029/91GL02788

Hansen, J., M. K. I. Sato, R. Ruedy, A. Lacis, K. Asamoah, S. Borenstein, and H. Wilson, 1996: A Pinatubo climate modeling investigation. In The Mount Pinatubo Eruption: Effects on the Atmosphere and Climate, pp. 233–272, Springer, Berlin, Heidelberg.

Harrop, B. E., & Hartmann, D. L. (2016). The role of cloud radiative heating in determining the location of the ITCZ in aquaplanet simulations. Journal of Climate, 29(8), 2741–2763. https://doi.org/10.1175/JCLI-D-15-0521.1

Haslehner, K., Gasparini, B., & Voigt, A. (2024). Radiative heating of high-level clouds and its impacts on climate. Journal of Geophysical Research: Atmospheres, 129(12), e2024JD040850. https://doi.org/10.1029/2024JD040850

Hersbach, H., Bell, B., Berrisford, P., Hirahara, S., Horányi, A., Muñoz-Sabater, J., … Thépaut, J.-N. (2020). The ERA5 global reanalysis. Quarterly Journal of the Royal Meteorological Society, 146(730), 1999–2049. https://doi.org/10.1002/qj.3803

Iacono, M. J., J. S. Delamere, E. J. Mlawer, M. W. Shephard, S. A. Clough, and W. D. Collins, 2008: Radiative forcing by long-lived greenhouse gases: Calculations with the AER radiative transfer models. *J. Geophys. Res.*, 113, D13103, https://doi.org/10.1029/2008JD009944

IPCC, 2021: Technical Summary. In *Climate Change 2021: The Physical Science Basis. Contribution of Working Group I to the Sixth Assessment Report of the Intergovernmental Panel on Climate Change*, V. Masson-Delmotte, P. Zhai, A. Pirani, S. L. Connors, C. Péan, S. Berger, and coeditors, Eds., Cambridge University Press, Cambridge, United Kingdom and New York, NY, USA, 33–144, https://doi.org/10.1017/9781009157896.002

Kiehl, J. T., and K. E. Trenberth, 1997: Earth's annual global mean energy budget. Bull. Amer. Meteor. Soc., 78, 197–208, https://doi.org/10.1175/1520-0477(1997)078<0197:EAGMEB>2.0.CO;2

Khairoutdinov, M. F., and D. A. Randall, 2003: Cloud resolving modeling of the ARM summer 1997 IOP: Model formulation, results, uncertainties, and sensitivities. J. Atmos. Sci., 60, 607–625, https://doi.org/10.1175/1520-0469(2003)060<0607:CRMOTA>2.0.CO;2

Laakso, A., D. Visioni, U. Niemeier, S. Tilmes, and H. Kokkola, 2024: Dependency of the impacts of geoengineering on the stratospheric sulfur injection strategy. Part 2: How changes in the hydrological cycle depend on the injection rate and model used. *Earth Syst. Dynam.*, 15, 405–427, https://doi.org/10.5194/esd-15-405-2024

Lacis, A. A., 2015: Volcanic aerosol radiative properties. Past Global Changes Mag., 23(2), 50–51, https://doi.org/10.22498/pages.23.2.50

McGraw, Z., and L. M. Polvani, 2024: How volcanic aerosols globally inhibit precipitation. *Geophys. Res. Lett.*, 51, e2023GL107930, https://doi.org/10.1029/2023GL107930

McGraw, Z., K. DallaSanta, L. M. Polvani, K. Tsigaridis, C. Orbe, and S. E. Bauer, 2024: Severe global cooling after volcanic super-eruptions? The answer hinges on unknown aerosol size. *J. Clim.*, 37, 1449–1464, https://doi.org/10.1175/JCLI-D-23-0282.1

McGraw, Z., L. M. Polvani, B. Gasparini, E. K. Van de Koot, and A. Voigt, 2025: The cloud radiative response to surface warming weakens hydrological sensitivity. Geophys. Res. Lett., 52, e2024GL112368, https://doi.org/10.1029/2024GL112368

Morrison, H., and J. A. Milbrandt, 2015: Parameterization of cloud microphysics based on the prediction of bulk ice particle properties. Part I: Scheme description and idealized tests. J. Atmos. Sci., 72, 287–311, https://doi.org/10.1175/JAS-D-14-0065.1

Muller, C. J., and P. A. O'Gorman, 2011: An energetic perspective on the regional response of precipitation to climate change. Nat. Climate Change, 1, 266–271, https://doi.org/10.1038/nclimate1169

National Academies of Sciences, Engineering, and Medicine, 2021: *Reflecting Sunlight: Recommendations for Solar Geoengineering Research and Research Governance*. National Academies Press, Washington, DC, 246 pp., https://doi.org/10.17226/25762

O'Gorman, P. A., R. P. Allan, M. P. Byrne, and M. Previdi, 2012: Energetic constraints on precipitation under climate change. *Surv. Geophys.*, 33, 585–608, https://doi.org/10.1007/s10712-011-9159-6

Orbe, C., D. Rind, J. Jonas, L. Nazarenko, G. Faluvegi, L. T. Murray, and G. A. Schmidt, 2020: GISS ModelE2.2: A climate model optimized for the middle atmosphere—2. Validation of large-scale transport and evaluation of climate response. *J. Geophys. Res. Atmos.*, 125, e2020JD033151, https://doi.org/10.1029/2020JD033151

Raiter, D., Z. McGraw, and L.M. Polvani, L.M., 2025: Non-linear and distinct responses of temperature and precipitation to volcanic eruptions with stratospheric sulfur injection from 5 to 160 Tg. Geophys. Res. Lett. 52(20), e2025GL116372, https://doi.org/10.1029/2025GL116372

Simpson, I. R., S. Tilmes, J. H. Richter, B. Kravitz, D. G. MacMartin, M. J. Mills, and A. G. Pendergrass, 2019: The regional hydroclimate response to stratospheric sulfate geoengineering and the role of stratospheric heating. *J. Geophys. Res. Atmos.*, 124, 12,587–12,616, https://doi.org/10.1029/2019JD031093

Stubenrauch, C. J., G. Caria, S. E. Protopapadaki, and F. Hemmer, 2021: 3D radiative heating of tropical upper tropospheric cloud systems derived from synergistic A-Train observations and machine learning. Atmos. Chem. Phys., 21, 1015–1034, https://doi.org/10.5194/acp-21-1015-2021

Visioni, D., E. M. Bednarz, W. R. Lee, B. Kravitz, A. Jones, J. M. Haywood, and D. G. MacMartin, 2023: Climate response to off-equatorial stratospheric sulfur injections in three Earth system models—Part 1: Experimental protocols and surface changes. *Atmos. Chem. Phys.*, 23, 663–685, https://doi.org/10.5194/acp-23-663-2023

Visioni, D., D. G. MacMartin, B. Kravitz, O. Boucher, A. Jones, T. Lurton, and S. Tilmes, 2021: Identifying the sources of uncertainty in climate model simulations of solar radiation modification with the G6sulfur and G6solar GeoMIP simulations. *Atmos. Chem. Phys.*, 21, 10,039–10,063, https://doi.org/10.5194/acp-21-10039-2021

Voigt, A., S. North, B. Gasparini, and S.-H. Ham, 2024: Atmospheric cloud-radiative heating in CMIP6 and observations and its response to surface warming. Atmos. Chem. Phys., 24, 9749–9775, https://doi.org/10.5194/acp-24-9749-2024

Weisenstein, D. K., D. Visioni, H. Franke, U. Niemeier, S. Vattioni, G. Chiodo, and D. W. Keith, 2022: An interactive stratospheric aerosol model intercomparison of solar geoengineering by stratospheric injection of $SO_2$ or accumulation-mode sulfuric acid aerosols. Atmos. Chem. Phys., 22, 2955–2973, https://doi.org/10.5194/acp-22-2955-2022

Wing, A. A., K. A. Reed, M. Satoh, B. Stevens, S. Bony, and T. Ohno, 2018a: Radiative–convective equilibrium model intercomparison project. *Geosci. Model Dev.*, **11**, 793–813, https://doi.org/10.5194/gmd-11-793-2018

Wing, A. A., K. Emanuel, C. E. Holloway, and C. Muller, 2018b: Convective self-aggregation in numerical simulations: A review. In *Shallow Clouds, Water Vapor, Circulation, and Climate Sensitivity*, 1–25, Springer, https://doi.org/10.1007/978-3-319-77273-8_1

Wing, A. A., L. G. Silvers, and K. A. Reed, 2024: RCEMIP-II: mock-Walker simulations as phase II of the Radiative-Convective Equilibrium Model Intercomparison Project. *Geosci. Model Dev.*, 17, 6195–6225, https://doi.org/10.5194/gmd-17-6195-2024

Zhang, Y., Z. Jin, and M. Sikand, 2021: The top-of-atmosphere, surface and atmospheric cloud radiative kernels based on ISCCP-H datasets: Method and evaluation. J. Geophys. Res. Atmos., 126, e2021JD035053, https://doi.org/10.1029/2021JD035053